\documentclass[11pt]{article} 
\usepackage{graphicx,amssymb,mathrsfs,array,multirow}  
\begin{document}  
%\pagestyle{empty}  
%\begin{flushright}  
%\texttt{hep-ph/yymmnnn}\\  
%\end{flushright}  
 
\vskip 30pt

%opening
\begin{center}  
{\Large{\bf 
Scotogenic S3 symmetric generation of realistic neutrino mixing}}\\
\vspace*{1cm}  
\renewcommand{\thefootnote}{\fnsymbol{footnote}}  
{ {\sf Soumita Pramanick$^{1}$\footnote{email: soumita509@gmail.com}}
} \\  
\vspace{10pt}  
{\small  {\em $^1$Harish-Chandra Research Institute, Chhatnag Road, Jhunsi,
Allahabad 211019, India \\
}}
\normalsize  

\normalsize

\end{center}  

%%%%%%%%%%%%%%%%%%%%%%%%%%%%%%%%%%%% Abstract %%%%%%%%%%%%%%%%%%%%%%%%%%%%%%%%%%%%%%%%%%
\begin{abstract} 
\textit{
Realistic neutrino mixing is achieved at one-loop level
radiatively using $S3\times Z_2$ symmetry.
The model comprises of two right-handed neutrinos, maximally mixed
to produce the structure of the left-handed Majorana neutrino mass matrix 
characterized by $\theta_{13}=0$, $\theta_{23}=\pi/4$ and any value of $\theta_{12}^0$ particular
to the Tribimaximal (TBM),
Bimaximal (BM) and Golden Ratio (GR) or other mixings. A small deviation from this maximal mixing between the 
two right-handed neutrinos could generate non-zero $\theta_{13}$, shifts of the atmospheric mixing
angle $\theta_{23}$ from $\pi/4$ and also could correct the solar mixing angle $\theta_{12}$
by a small amount altogether in a single step.
In this scotogenic mechanism of generating non-zero $\theta_{13}$ by shifting from maximal mixing in the
right-handed neutrino sector, two $Z_2$ odd inert scalar $SU(2)_L$ doublets were used, the 
lightest of which can serve as a dark matter candidate.
}

\end{abstract}  

\renewcommand{\thesection}{\Roman{section}} 
\setcounter{footnote}{0} 
\renewcommand{\thefootnote}{\arabic{footnote}} 
\noindent

%%%%%%%%%%%%%%%%%%%%%%%%%%%%%%%%%%%% Introduction %%%%%%%%%%%%%%%%%%%%%%%%%%%%%%%%%%%%%%%%%%

\section{Introduction}
Neutrinos oscillate owing to their massive nature as established by the oscillation experiments.
The mass eigenstates and flavour eigenstates are different and are related by the Pontecorvo, Maki, Nakagawa, Sakata --
PMNS -- matrix:
\begin{eqnarray}
U = \left(
          \begin{array}{ccc}
          c_{12}c_{13} & s_{12}c_{13} & -s_{13}e^{-i\delta}  \\
 -c_{23}s_{12} + s_{23}s_{13}c_{12}e^{i\delta} & c_{23}c_{12} +
s_{23}s_{13}s_{12}e^{i\delta}&  -s_{23}c_{13}\\
 - s_{23}s_{12} - c_{23}s_{13}c_{12}e^{i\delta}& - s_{23}c_{12} +
c_{23}s_{13}s_{12}e^{i\delta} & c_{23}c_{13} \end{array} \right)
\;\;.
%\nonumber
\label{PMNS}
\end{eqnarray}
Here $c_{ij} = \cos \theta_{ij}$ and $s_{ij} = \sin \theta_{ij}$. 
Needless to mention that the mass eigenstates are non-degenerate.

Non-zero $\theta_{13}$, though small in comparison to the other mixing
angles was discovered in 2012 by the short-baseline reactor anti-neutrino 
experiments \cite{t13_s3}. Before these non-zero $\theta_{13}$ results,
models were studied in literature that correspond to Tribimaximal (TBM),
Bimaximal (BM) and Golden Ratio (GR) mixings (that we now onwards collectively refer
as popular lepton mixings).
All these mixings have $\theta_{13}=0$, $\theta_{23}=\pi/4$
and tuning $\theta_{12}^0$ to the specific values as shown in Table \ref{t1}
produced the different mixing patterns viz. TBM, BM and GR.

Setting $\theta_{13}=0$ and $\theta_{23}=\pi/4$ in
Eq. (\ref{PMNS}) will yield a general structure for all popular mixing as:
\begin{equation}
U^0=
\pmatrix{\cos \theta_{12}^0 & \sin \theta_{12}^0  & 0 \cr -\frac{\sin
\theta_{12}^0}{\sqrt{2}} & \frac{\cos \theta_{12}^0}{\sqrt{2}} &
-{1\over\sqrt{2}} \cr
-\frac{\sin \theta_{12}^0}{\sqrt{2}} & \frac{\cos
\theta_{12}^0}{\sqrt{2}}  & {1\over\sqrt{2}}}.
\label{mix0}
\end{equation}

%--------------------------- 
\begin{table}[tb]
\begin{center}
\begin{tabular}{|c|c|c|c|}
\hline
  Model &TBM &BM & GR \\ \hline
$\theta^0_{12}$ & 35.3$^\circ$ & 45.0$^\circ$  & 31.7$^\circ$ \\ \hline
\end{tabular}
\end{center}
\caption{\sf{The values $\theta^0_{12}$ corresponding to various popular lepton mixings
namely, TBM, BM, and GR  patterns.}}
\label{t1}
\end{table}
%---------------------

The current 3$\sigma$ global fit \cite{Gonzalez_s3, Valle_s3} for $\theta_{13}$, $\theta_{23}$ and $\theta_{12}$ as from NuFIT3.2 of 2018 \cite{Gonzalez_s3} are:
\begin{eqnarray}
\theta_{12} &=&(31.42 - 36.05)^\circ, \nonumber \\
\theta_{23} &=& (40.3-51.5)^\circ \,, \nonumber \\
\theta_{13} &=& (8.09 - 8.98)^\circ.
\label{results}
\end{eqnarray}

So popular mixing and non-zero $\theta_{13}$ observations
are not in harmony. Several model-building exercises have been 
taking place since the observation of non-zero $\theta_{13}$  to
include it in the popular mixing framework.
In \cite{br}, the possibility of smallness of $\theta_{13}$ and $\Delta m^2_{solar}$ 
to have a common origin was explored.
In some efforts \cite{pr} a dominant component was characterized by larger oscillation parameters such as $\Delta m^2_{atmos}$ and $\theta_{23} = \pi/4$, whereas the smaller mixing parameters viz. non-zero
$\theta_{13},$ $\theta_{12}$, solar splitting and deviation of atmospheric mixing from maximality
were produced by a smaller see-saw \cite{seesaw} component as perturbation to the 
dominant one\footnote{For some earlier models with similar goals, see \cite{old}.}.
In \cite{models, LuhnKing} the mixing angle $\theta_{13}=0$ was produced using 
various symmetries and non-vanishing $\theta_{13}$ was produced
by perturbation to these symmetric forms.   

The popular mixings were amended at tree-level using a two-component Lagrangian
with discrete symmetries $A4$, $S3$ in \cite{ourS3,newA4}. In these models, 
type II see-saw yielded the dominant component that gave the popular mixing, corrections
to which were offered by type I see-saw sub-dominant component. 
Similar enterprise just for the no solar mixing (NSM) 
case i.e., $\theta_{12}^0=0$ using $A4$ was pursued\footnote{
The dominant type II seesaw had vanishing solar splitting, thus
one can make use of degenerate perturbation theory 
to get large solar mixing.} in \cite{ourA4}.
In \cite{Ma_rad} TBM was obtained radiatively using $A4$. 
Recent works with realistic neutrino mixings can be found in 
\cite{newMa, tanimoto}.

Here we discuss a radiative $S3\times Z_2$ model\footnote{
A brief account on discrete group $S3$ in presented in Appendix \ref{GroupS3} of the paper.}.
Some earlier works on $S3$ in context of neutrino mass are \cite{S3older, S3ma}.
Neutrino mass with  $S3\times Z_2$ within left-right symmetry
 was studied in \cite{LRSMs3}.
A common practice \cite{Idemo} was to find a symmetry among the
three neutrinos that can produce a mass matrix that can be expressed as
a linear combination of a democratic matrix $M_{dem}$ and 
an identity matrix $I$, like $c_1 I + c_2 M_{dem}$ with 
$c_1$ and $c_2$ being two complex numbers.
This could serve as a reasonable
scenario to start with from which 
some models obtained realistic mixing through perturbation to such 
initial structures \cite{Idemo} whereas in some models \cite{Idem2}
various GUT symmetries or extra-dimensional theories were considered
to generate these initial structures and renormalization 
group effects at high energies were explored to obtain realistic mixing.
Another way \cite{S3other} of constructing $S3$ models is to have a 
3-3-1 local gauge symmetry, and later on associate it to a $(B-L)$ extension
or use soft breaking of $S3$. Since $S3$ has irreducible representations of 
one-dimension and two-dimension, the latter can be used to obtain 
maximal mixing in the $\nu_\mu-\nu_\tau$ block \cite{S3mutau}.
Collider signatures of $S3$ flavour symmetry was vividly studied in \cite{gb}.
$S3$ models are also studied in quark sector \cite{S3quarks}. Some earlier
studies on scotogenic models can be found in \cite{scotogenic_Prof_Valle}.

In this work our objective is to use $S3$ to radiatively\footnote{
A systematic analysis of radiative neutrino mass models can be found in \cite{radreview}.} obtain:
\begin{enumerate}
\item The structure of the mixing matrix of popular mixing kind as shown in Eq. (\ref{mix0}) that is characterized by $\theta_{13}=0$, $\theta_{23}=\pi/4$ and $\theta_{12}^0$ of any of the alternatives displayed in Table \ref{t1}.
\item  Realistic neutrino mixings i.e.,  precisely non-zero $\theta_{13}$, shifts of atmospheric mixing angle $\theta_{23}$ from maximality and tiny corrections to the solar mixing angle $\theta_{12}$.
\end{enumerate}
In this radiative $S3\times Z_2$ model, neutrino masses and mixings are generated at one-loop.
The model has two right-handed neutrinos comprising an $S3$ doublet, that are maximally mixed to 
obtain the structure as required by popular mixings as in Eq. (\ref{mix0}).
A small deviation from this maximal mixing in the right-handed neutrino sector could 
produce in a single step non-zero $\theta_{13}$, shifts of $\theta_{23}$ from $\pi/4$ and small
corrections to $\theta_{12}$ as is required by the mixing to be realistic.
To achieve this, two $Z_2$ odd scalars $\eta_i$, ($i=1,2$), were required, the lightest
among them can be a good dark matter candidate.
A similar analysis based on $A4$ was performed where instead of using
deviations from maximal mixing between the two right-handed neutrino states 
to generate non-zero $\theta_{13}$, small mass splittings between two right-handed neutrinos
were used in \cite{radA4}.

\section{The $S3\times Z_2$ Model}
In mass basis the left-handed neutrino Majorana mass matrix
 is $M^{mass}_{\nu L}=$ $diag\; (m_1, m_2, m_3)$. One can transport this in 
its flavour basis with help of the common form of the popular
lepton mixing matrix $U^0$ in Eq. (\ref{mix0}) as:
\begin{equation}
M^{flavour}_{\nu L}=U^0 M^{mass}_{\nu L} U^{0T}=\pmatrix{a & c & c \cr
c & b & d \cr c & d & b}.
\label{abc}
\end{equation}
The $a,b,c$ and $d$ used here are given by:
\begin{eqnarray}
a&=& m_1\cos^2 \theta_{12}^0+m_2\sin^2 \theta_{12}^0\nonumber\\
b&=&\frac{1}{2}\left(m_1\sin^2 \theta_{12}^0+m_2\cos^2 \theta_{12}^0+m_3\right)\nonumber\\
c&=&\frac{1}{2\sqrt{2}}\sin 2\theta_{12}^0 (m_2-m_1)\nonumber\\
d&=&\frac{1}{2}\left(m_1\sin^2 \theta_{12}^0+m_2\cos^2 \theta_{12}^0-m_3\right).
\label{abcd}
\end{eqnarray}
Thus,
\begin{equation}
\tan 2\theta_{12}^0=\frac{2\sqrt 2 c}{b+d-a}.
\end{equation}
It is essential for $a,b,c$ and $d$ to be non-zero for the neutrino masses to be realistic
and non-degenerate.

Our prime intent is to generate the form of $M^{flavour}_{\nu L}$ in Eq. (\ref{abc}) 
radiatively with one-loop.
Thus one has to designate each of the fields in our model with particular $S3\times Z_2$
quantum numbers. There are two right-handed neutrinos present in the model. Maximal mixing 
between these two right-handed neutrino fields can produce the desired form of left-handed 
Majorana neutrino mass matrix in Eq. (\ref{abc}) that corresponds to $\theta_{13}=0$,
$\theta_{23}=\pi/4$ and $\theta_{12}^0$ of the popular lepton mixing scenarios.
After obtaining the form in Eq. (\ref{abc}), we will see in due course, a slight shift from this 
maximal mixing between the right-handed neutrino states is capable of yielding realistic neutrino mixings,
viz. non-zero $\theta_{13}$, deviation of atmospheric mixing $\theta_{23}$ from $\pi/4$ as well as
small corrections to solar mixing $\theta_{12}$.

\begin{figure}[tbh]
\begin{center}
\includegraphics[scale=0.22,angle=0]{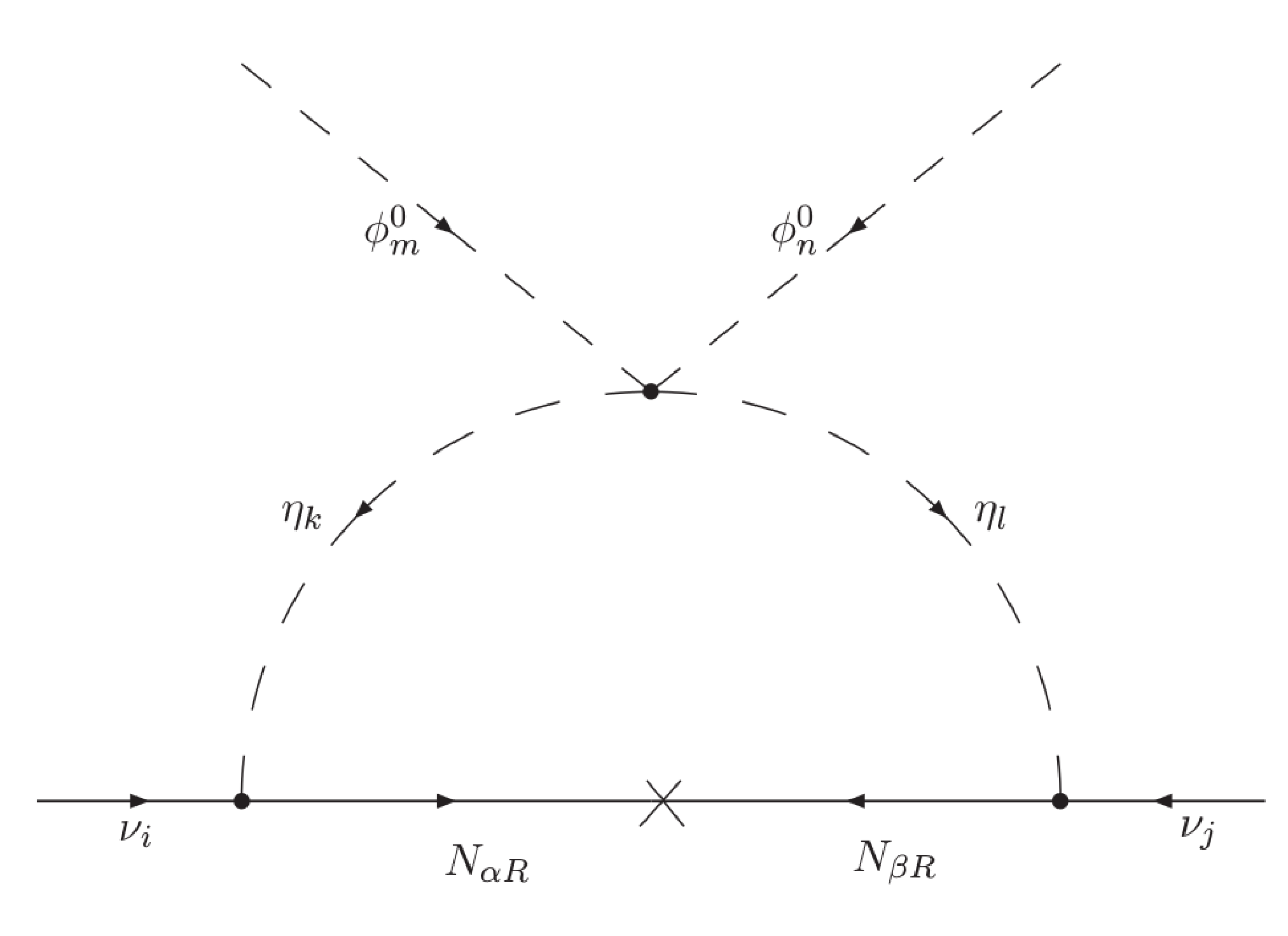}
\end{center}
\caption{\sf  One-loop scotogenic neutrino mass generation using $S3\times Z_2$
symmetry.
}
\label{radfig} 
\end{figure}

The model has the three left-handed lepton $SU(2)_L$ 
doublets $L_{\zeta_L}\equiv(\nu_\zeta \ \ \zeta^-)_L^T$
where $\zeta=e,\mu,\tau$, out of which $L_{\mu_L}$ and $L_{\tau_L}$ 
comprise a doublet of $S3$ whereas $L_{e L}$
remains a singlet under $S3$. Apart from these there are 
two Standard Model (SM) gauge singlet right-handed neutrinos $N_{\alpha R}$,
($\alpha=1,2$) that transform as a doublet under $S3$. 
The scalar spectrum of the model has a couple of 
inert $SU(2)_L$ doublet scalars, 
$\eta_i\equiv(\eta_i^+, \eta_i^0)^T$, $(i=1,2)$, 
forming an $S3$ doublet ($\eta$). We also 
have two other $SU(2)_L$ doublet scalars, 
namely $\Phi_j\equiv(\phi_j^+, \phi_j^0)^T$, $(j=1,2)$, 
that are combined to form an $S3$ doublet ($\Phi$).
Besides the $S3$, the model also has an unbroken $Z_2$ symmetry
under which all other fields except the right-handed 
neutrinos and the scalar $\eta$ are even. After 
spontaneous symmetry breaking (SSB), $\phi_j$ get
vacuum expectation value (vev), but $\eta_i$ do not. 
Let $v_j$ be the vevs of $\phi_j^0$ i.e., $\langle \Phi_j\rangle\equiv v_j$, 
$(j=1,2)$.
Fields and their specific charges are shown in Table \ref{fields}.
We deal with the neutrino sector only in this model.
The charged lepton mass matrix is diagonal in the basis in which we perform the 
analysis and the entire mixing comes from the neutrino sector.
%--------------------------- 
\begin{table}[tb]
\begin{center}
\begin{tabular}{|c|c|c|c|}
\hline
\sf{Leptons} & $SU(2)_L$ & $S3$ & $Z_2$ \\ \hline
 & & &  \\ 
$L_{e_L}\equiv\pmatrix{\nu_e& e^-}_L$ & $2$ & $1$ & $1$ \\
 & & &  \\ 
 \hline
 & & &  \\ 
$L_{\zeta_L}\equiv\pmatrix{
\nu_\mu & \mu^- \cr \nu_\tau & \tau^- }_L$& $2$ & $2$ & $1$ \\
 & & &  \\ 
 \hline
 & & &  \\ 
$N_{\alpha R}\equiv \pmatrix{N_{1R}\cr
N_{2R}} $ & $1$ & $2$ & $-1$ \\ 
 & & &  \\ 
\hline
\hline
\sf{Scalars}& $SU(2)_L$ & $S3$ & $Z_2$ \\ \hline
 & & &  \\ 
$\Phi \equiv \pmatrix{\phi_1^+ & \phi_1^0\cr
\phi_2^+ & \phi_2^0 }$ & $2$ & $2$ & $1$ \\ 
 & & &  \\ 
\hline
 & & &  \\ 
$\eta\equiv \pmatrix{\eta_1^+ & \eta_1^0\cr
\eta_2^+ & \eta_2^0 }$ & $2$ & $2$ & $-1$ \\ 
 & & &  \\ 
\hline
\end{tabular}
\end{center}
\caption{\sf{All fields along with their respective charges.
We confine this model to neutrino sector only.}}
\label{fields}
\end{table}
%---------------------
\vskip 2pt
Neutrino mass can be generated radiatively at one-loop level
from Fig. \ref{radfig}. The neutrino mass matrix will receive 
contributions from the following terms of the $S3\times Z_2$ invariant scalar potential from the
scalar four-point vertex\footnote{
Two $\eta$ are created and two $\phi$ are destroyed at the scalar four point
vertex causing terms of $(\eta^\dagger \phi)(\eta^\dagger \phi)$ nature to be
pertinent among other terms in the scalar potential. The complete scalar
potential containing all the terms can be found in Appendix \ref{potential}.
}: 
\begin{eqnarray}
V_{relevant}&\supset& \lambda_1 \left[ \left\{(\eta_2^\dagger \phi_2+\eta_1^\dagger \phi_1)^2 \right  \}  + h.c.\right]+ \lambda_2\left[ \left\{(\eta_2^\dagger \phi_2- \eta_1^\dagger \phi_1 )^2 \right \} +h.c.
\right]\nonumber\\
&+&\lambda_3\left[ \left\{(\eta_1^\dagger \phi_2)(\eta_2^\dagger \phi_1)+(\eta_2^\dagger \phi_1)(\eta_1^\dagger \phi_2)\right  \} 
+h.c.\right].
\label{potential}
\end{eqnarray}
Here all the quartic couplings $\lambda_j$ ($j=1,2,3$) are taken real.
%---------------------------------------------------------------------------
%---------------------------------------------------------------------------

At all the three vertices of Fig. \ref{radfig}, all symmetries 
are conserved. 
The Dirac vertices conserving $S3\times Z_2$ can
be written as:
\begin{equation}
\mathscr{L}_{Yukawa}= y_1\left[(\overline{N}_{2R}\eta_2^0+\overline{N}_{1R}\eta_1^0)\nu_e\right]
+y_2\left[(\overline{N}_{1R}\eta_2^0)\nu_\tau+ (\overline{N}_{2R}\eta_1^0)\nu_\mu\right]+h.c.
\label{yukawa}
\end{equation}
Since the left-handed neutrinos ${\nu_\zeta}_L$ transform as
doublet of $S3$ for $(\zeta=\mu,\tau)$ and invariant under $S3$ if $\zeta=e$,
the Yukawa couplings involved are different for $(\zeta=\mu,\tau)$ and $\zeta=e$,
namely, $y_1$ for $\zeta=e$ and $y_2$ for $(\zeta=\mu,\tau)$ respectively.

Let us now have a look at the right-handed neutrino sector.
Recall we have two SM gauge singlet right-handed neutrinos, $N_{1R}$ and $N_{2R}$,
that transform as a doublet of $S3$.
Thus the $S3\times Z_2$ invariant direct mass term
for the right-handed neutrinos will look like:
\begin{equation}
\mathscr{L}_{right-handed \, neutrinos}= \frac{1}{2} m_{R_{12}}
\left[N_{1R}^T C^{-1}N_{2R}+N_{2R}^T C^{-1}N_{1R}\right].
\end{equation}
Thus $S_3$ symmetry allows a symmetric mass matrix with only
non-zero off-diagonal terms for the right-handed neutrinos.
If one allows soft breaking of $S3$ at the scale where right-handed
neutrinos get mass by introducing terms like:
\begin{equation}
\mathscr{L}_{soft}= \frac{1}{2}\left[ m_{R_{11}}
N_{1R}^T C^{-1}N_{1R}+m_{R_{22}}N_{2R}^T C^{-1}N_{2R}\right]
\end{equation}
to get non-zero diagonal entries, then one can write the right-handed
neutrino mass matrix as:
\begin{equation}
M_{\nu_R}=\frac{1}{2}\pmatrix{m_{R_{11}} & m_{R_{12}}\cr
m_{R_{12}} & m_{R_{22}}
}.
\label{mrh}
\end{equation}
The symmetric structure of the matrix in Eq. (\ref{mrh}) also reflects its Majorana nature.  

Before moving on, let us have a brief discussion
about the dark matter candidates in the model.
It is a common practice in literature to stabilize 
dark matter candidate with discrete symmetries 
like $Z_2$. Thus the $Z_2$ symmetry is an indication that this
model can provide dark matter candidate. Both the right-handed neutrinos
and the scalar fields $\eta$ are odd under $Z_2$,
among which $\eta$ are chosen lighter than the 
right-handed neutrinos $N_{\alpha R}$,
($\alpha=1,2$). Although from the $m^2_\eta$
term in Eq. (\ref{totalpotential}), the $\eta_i$, ($i=1,2$)
appear to be degenerate in mass, since the $S3$ symmetry
is softly broken in the right-handed neutrino sector, it can lead
to small mass splitting between the two $\eta_i$, ($i=1,2$).
The lightest among the two $\eta_i$, ($i=1,2$) can be the 
dark matter candidate.
%-------------------

With the model ingredients ready, at this stage, we are in a
position to present a basic description of the left-handed Majorana 
neutrino mass matrix arising from Fig. \ref{radfig}, the detailed expressions
for which will be provided at a later stage of our analysis.
To set the stage of the discussion, let us first sketchily indicate how the
elements of the left-handed neutrino mass matrix will receive contributions from
this one-loop diagram \cite{Ma_loop} in Fig. \ref{radfig}. 
Let us make a few simplifying assumptions to make the expressions look
less complicated at the moment. For this purpose, let $\lambda$ commonly represent 
some combinations of  the three quartic couplings given in Eq. (\ref{potential}) 
i.e., $\lambda_1$, $\lambda_2$ and $\lambda_3$.
Also the splitting between the masses of $\eta_1$ and $\eta_2$ comprising the $S3$ doublet
is neglected and $m_0$ is assumed to be the common mass of them.
Further, if the real part of $\eta_j^0$ is denoted by $\eta_{Rj}$ and $\eta_{Ij}$ be the 
imaginary part of $\eta_j^0$, then difference between the masses of $\eta_{Rj}$ and $\eta_{Ij}$
can be taken proportional to $\lambda v_j$ and can be small in general.

It is imperative to note that under $S3$, $\nu_e$ is invariant whereas 
$\nu_\zeta$ ($\zeta=\mu,\tau$) transform as doublet. This feature will manifest through the 
Yukawa couplings (see Eq. (\ref{yukawa})) at the two Dirac vertices which in its turn will
dictate the structure of the left-handed neutrino mass matrix.
Let $z\equiv \frac{m_R^2}{m_0^2}$, where
$m_R$ is the average mass of the heavy right-handed neutrino states.
Since $z$ always appears only in the logarithm
we do not distinguish between the masses of the different right-handed neutrinos
for the purpose of defining $z$ throughout. 
Under this assumption the second diagonal entry, for example, will have the form, 
\begin{equation}
(M_{\nu_L}^{flavour})_{22}=\lambda \frac{v_m v_n}{8 \pi^2}
\frac{y_2^2 }{m_{R_{22}}} 
\left[ \ln z -1 \right].
\label{m22}
\end{equation}
It is noteworthy that Eq. (\ref{m22})
is valid in the limit $m_R^2>>m_0^2$.
For $(M_{\nu_L}^{flavour})_{22}$, as noted earlier in Eq. (\ref{yukawa}), $\nu_\mu$
couples only to $N_{2R}$, thus at both the Dirac vertices $N_{2R}$
will couple with $\nu_\mu$. Hence the ($2,2$) element of the left handed
neutrino mass matrix will get contribution from $m_{R_{22}}$ only. Also
$y_2$ is the only Yukawa coupling that will appear since we are
dealing with $\nu_\mu$ at both the Dirac vertices for $(M_{\nu_L}^{flavour})_{22}$.
From similar arguments, one can obtain expression 
for $(M_{\nu_L}^{flavour})_{33}$ just by replacing $m_{R_{22}}$ by $m_{R_{11}}$
in Eq. (\ref{m22}). 

Let us now concentrate on the off-diagonal ($2,3$) entry.
Thus one has to consider $\nu_\mu$ at one of the Dirac vertices
and $\nu_\tau$ at the other.
From Eq. (\ref{yukawa}), one can note that $\nu_\mu$ couples to
$N_{2R}$ only whereas $\nu_\tau$ does so with $N_{1R}$. 
Thus at one of the Dirac vertices we will have $N_{1R}$ and $N_{2R}$
at the other. Therefore, off-diagonal entries from right-handed neutrino
mass matrix will come into play and
$(M_{\nu_L}^{flavour})_{23}$ will get 
contributions from $m_{R_{12}}$ in addition to that from $m_{R_{11}}$
and $m_{R_{22}}$. Needless to mention that the Yukawa coupling involved
will be $y_2$ as can be seen from Eq. (\ref{yukawa}).
Thus one can write,
\begin{equation}
(M_{\nu_L}^{flavour})_{23}=\lambda \frac{v_m v_n}{8 \pi^2}
\frac{y_2^2 m_{R_{12}}}{m_{R_{11}}m_{R_{22}}} 
\left[ \ln z -1 \right].
\label{m23}
\end{equation}
While writing down Eq. (\ref{m23}) we are taking into account
the mass insertion approximation.
In similar spirit, one can write down expressions for ($1,1$), ($1,2$)
and the ($1,3$) entries of the left-handed Majorana neutrino mass matrix.

For notational ease, let us absorb
everything else present in the RHS of expressions
for the elements of the left-handed Majorana neutrino mass
matrix as in Eq. (\ref{m22}) and Eq. (\ref{m23}) except the 
Yukawa couplings, quartic couplings and the vevs in loop contributing factors 
say $r_{\alpha\beta}$ given by:
\begin{eqnarray}
r_{11}&\equiv&\frac{1}{8 \pi^2 m_{R_{11}}}\left[ \ln z -1 \right],\nonumber\\
r_{22}&\equiv&\frac{1}{8 \pi^2 m_{R_{22}}}\left[ \ln z -1 \right],\nonumber\\
r_{12}&\equiv&\frac{m_{R_{12}}}{ 8 \pi^2 m_{R_{11}}m_{R_{22}}}\left[ \ln z -1 \right].
\label{rij}
\end{eqnarray}

%-----------------------

From Eqs. (\ref{m22}), (\ref{m23}), (\ref{rij}) and (\ref{potential}), the 
left-handed neutrino Majorana mass matrix radiatively
generated at one-loop as shown in Fig. \ref{radfig} is:
\begin{equation}
M_{\nu_L}^{flavour}=\pmatrix{ \chi_1
& \chi_4 & \chi_5 \cr
\chi_4
& \chi_2 & \chi_6 \cr 
\chi_5
& \chi_6 & \chi_3
}
\label{mgeneral}
\end{equation}
where,
\begin{eqnarray}
\chi_1&\equiv&y_1^2\left[
4 r_{12} v_1v_2(\lambda_3+\lambda_1-\lambda_2)
+(r_{11}v_1^2+r_{22}v_2^2)(\lambda_1+\lambda_2)
\right]\nonumber\\
\chi_2&\equiv&y_2^2\left[r_{22}(\lambda_1+\lambda_2)v_1^2\right]\nonumber\\
\chi_3&\equiv&y_2^2\left[r_{11}(\lambda_1+\lambda_2)v_2^2\right]\nonumber\\
\chi_4&\equiv& y_1y_2\left[
r_{12}(\lambda_1+\lambda_2)v_1^2+2r_{22}(\lambda_3+\lambda_1-\lambda_2)v_1v_2
\right]\nonumber\\
\chi_5&\equiv& 
y_1y_2\left[
r_{12}(\lambda_1+\lambda_2)v_2^2+2r_{11}(\lambda_3+\lambda_1-\lambda_2)v_1v_2
\right]
\nonumber\\
\chi_6&\equiv& y_2^2\left[
2 r_{12}(\lambda_3+\lambda_1-\lambda_2)v_1v_2
\right].
\end{eqnarray}
Here $\langle \Phi_j\rangle\equiv v_j$ with ($j=1,2$).

For the left-handed neutrino mass matrix in Eq. (\ref{mgeneral})
to be of the form of Eq. (\ref{abc}) i.e., the structure needed
for $\theta_{13}=0$, $\theta_{23}=\pi/4$ and $\theta_{12}^0$ of the popular
mixing kind, we have to set $\chi_1\ne\chi_2=\chi_3$
as well as $\chi_4=\chi_5$. This is achieved when $v_1=v_2=v$ and
$r_{11}=r_{22}=r$. The condition $r_{11}=r_{22}=r$ when translated in terms of the
right-handed neutrino mass matrix in Eq. (\ref{mrh}) using Eq. (\ref{rij}) will lead to:
\begin{equation}
M_{\nu_R}=\frac{1}{2}\pmatrix{m_{R_{11}} & m_{R_{12}}\cr
m_{R_{12}} & m_{R_{11}}
}.
\label{mrh2}
\end{equation}
The matrix in Eq. (\ref{mrh2}) corresponds to maximal 
mixing in the right-handed neutrino sector.
Thus, to get the form 
of left-handed neutrino mass matrix as in 
Eq. (\ref{abc}) it is necessary to have $v_1=v_2=v$ as well as maximal mixing between
$N_{1R}$ and $N_{2R}$ i.e., we have to set $r_{11}=r_{22}=r$. Implementing these
constraints to the general form of the mass matrix in Eq. (\ref{mgeneral}) we get;
\begin{equation}
M_{\nu_L}^{flavour}=v^2\pmatrix{ y_1^2[4r_{12}\lambda_{123} +2r\lambda_{12}]
& y_1y_2[r_{12}\lambda_{12} +2r\lambda_{123}]
& y_1y_2[r_{12}\lambda_{12} +2r\lambda_{123}]\cr
y_1y_2[r_{12}\lambda_{12} +2r\lambda_{123}]
& y_2^2r\lambda_{12}& y_2^2(2r_{12}\lambda_{123})\cr 
y_1y_2[r_{12}\lambda_{12} +2r\lambda_{123}]
& y_2^2(2r_{12}\lambda_{123}) & y_2^2r\lambda_{12}
}.
\label{mchoicemaxmix}
\end{equation}
Here $\lambda_{12}\equiv\lambda_1+\lambda_2$ and $\lambda_{123}\equiv\lambda_3+\lambda_1-\lambda_2$.
To get the form of $M_{\nu_L}^{flavour}$ in Eq. (\ref{abc}), one has to identify:
\begin{eqnarray}
a&\equiv&y_1^2v^2[4r_{12}\lambda_{123} +2r\lambda_{12}]
= y_1^2v^2[4r_{12}(\lambda_3+\lambda_1-\lambda_2) +2r(\lambda_1+\lambda_2)]\nonumber\\
b&\equiv& y_2^2v^2r\lambda_{12}=y_2^2v^2r(\lambda_1+\lambda_2)\nonumber\\
c&\equiv&y_1y_2v^2[r_{12}\lambda_{12} +2r\lambda_{123}]=
y_1y_2v^2[r_{12}(\lambda_1+\lambda_2) +2r(\lambda_3+\lambda_1-\lambda_2)]
\nonumber\\
d&\equiv&y_2^2v^2(2r_{12}\lambda_{123})=y_2^2v^2[2r_{12}(\lambda_3+\lambda_1-\lambda_2)].
\label{id1}
\end{eqnarray} 
So far we are able to obtain the form of left-handed neutrino mass matrix required for 
$\theta_{13}=0$, $\theta_{23}=\pi/4$ and $\theta_{12}^0$ of the popular
mixing varieties. With this in hand, the obvious follow-up enterprise, as mentioned
earlier, will be to obtain realistic mixing viz. non-zero $\theta_{13}$, deviations 
of the atmospheric mixing angle $\theta_{23}$ from $\pi/4$ as well as tiny corrections to
$\theta_{12}$ also. To get such realistic neutrino mixing, we have to shift from
the choice of $r_{11}=r_{22}=r$, i.e., allow the two diagonal entries of the
right-handed neutrino mass matrix to slightly differ from each other.
In other words, let $r_{22}=r_{11}+\epsilon$, where $\epsilon$ is
a small quantity. 
Therefore, one gets back the general form of $M_{\nu R}$ in Eq. (\ref{mrh})
characterized by non-maximal mixing between $N_{1R}$ and $N_{2R}$.
Thus setting $r_{22}=r_{11}+\epsilon$ is precisely shifting from the
maximal mixing between the two right-handed neutrino states.
With $v_1=v_2=v$ still valid, we can get a dominant component
of $M_{\nu_L}^{flavour}$ as in Eq. (\ref{mchoicemaxmix}) denoted $M^0$ and a smaller contribution 
$M'$ proportional to $\epsilon$. Hence,
\begin{equation}
M_{\nu_L}^{flavour}=M^0+M'
\end{equation}   
with, 
\begin{equation}
M^0=
v^2\pmatrix{ y_1^2[4r_{12}\lambda_{123} +2r_{11}\lambda_{12}]
& y_1y_2[r_{12}\lambda_{12} +2r_{11}\lambda_{123}]
& y_1y_2[r_{12}\lambda_{12} +2r_{11}\lambda_{123}]\cr
y_1y_2[r_{12}\lambda_{12} +2r_{11}\lambda_{123}]
& y_2^2r_{11}\lambda_{12}& y_2^2(2r_{12}\lambda_{123})\cr 
y_1y_2[r_{12}\lambda_{12} +2r_{11}\lambda_{123}]
& y_2^2(2r_{12}\lambda_{123}) & y_2^2r_{11}\lambda_{12}
},
%\, {\rm and} \, 
\label{m0}
\end{equation} 
and
\begin{equation}
M'=\epsilon \pmatrix{ x
& y & 0\cr
y
& x' & 0\cr 
0
& 0 & 0
},
\end{equation} 
where,
\begin{eqnarray}
x&\equiv&y_1^2v^2\lambda_{12}
= y_1^2v^2(\lambda_1+\lambda_2)\nonumber\\
x'&\equiv& y_2^2v^2\lambda_{12}
= y_2^2v^2(\lambda_1+\lambda_2)\nonumber\\
y&\equiv&y_1y_2v^2\lambda_{123}=
y_1y_2v^2(\lambda_3+\lambda_1-\lambda_2).
\label{xy}
\end{eqnarray} 
$M^0$ in Eq. (\ref{m0}) will represent
the form of left-handed neutrino mass matrix
needed for $\theta_{13}=0$, $\theta_{23}=\pi/4$ and $\theta_{12}^0$ of the popular
mixing types as in Eq. (\ref{abc}) when we identify:
\begin{eqnarray}
a'&\equiv&y_1^2v^2[4r_{12}\lambda_{123} +2r_{11}\lambda_{12}]
= y_1^2v^2[4r_{12}(\lambda_3+\lambda_1-\lambda_2) +2r_{11}(\lambda_1+\lambda_2)]\nonumber\\
b'&\equiv& y_2^2v^2r_{11}\lambda_{12}=y_2^2v^2r_{11}(\lambda_1+\lambda_2)\nonumber\\
c'&\equiv&y_1y_2v^2[r_{12}\lambda_{12} +2r_{11}\lambda_{123}]=
y_1y_2v^2[r_{12}(\lambda_1+\lambda_2) +2r_{11}(\lambda_3+\lambda_1-\lambda_2)]
\nonumber\\
d'&\equiv&y_2^2v^2(2r_{12}\lambda_{123})=y_2^2v^2[2r_{12}(\lambda_3+\lambda_1-\lambda_2)]
\label{id2}
\end{eqnarray} 
in the same spirit\footnote{We are introducing the primed notation to differentiate from the
$r_{11}=r_{22}=r$ case.} as was done in case of Eq. (\ref{id1}).

With the help of non-degenerate perturbation theory we can calculate the 
corrections to eigenvalues and eigenvectors of $M^0$ from $M'$. The unperturbed flavour basis is given by the
columns of the mixing matrix $U^0$ as shown in Eq. (\ref{mix0}).
For ease of presentation it is useful to define,
\begin{equation}
 \gamma\equiv(b'-3d'-a') \ \  {\rm and} \ \ 
\rho\equiv\sqrt{a^{'2}+b^{'2}+8c^{'2}+d^{'2}-2a^{'}b^{'}-2a^{'}d^{'}+2b^{'}d^{'}}.
\label{grho}
\end{equation}
Thus the third ket after receiving first order corrections will take the form:
\begin{equation}
|\psi_3\rangle =
\pmatrix{\frac{\epsilon}{\gamma^2-\rho^2}
\left[\rho(\sqrt{2}y\cos 2\theta_{12}^0-x'\sin 2\theta_{12}^0)
-\gamma\sqrt{2}y\right]
\cr
-\frac{1}{\sqrt{2}}[1+\xi \epsilon]
\cr
\frac{1}{\sqrt{2}}[1-\xi \epsilon]}.
\label{ket3}
\end{equation}
Here, we have used
\begin{equation}
\xi\equiv[\gamma x' +\rho(x'\cos 2\theta_{12}^0+\sqrt{2}y\sin 2\theta_{12}^0)]/(\gamma^2-\rho^2).
\label{xi}
\end{equation}
If we consider CP-conserving scenario then,
\begin{equation}
\sin \theta_{13}=\frac{\epsilon}{\gamma^2-\rho^2}
\left[\rho(\sqrt{2}y\cos 2\theta_{12}^0-x'\sin 2\theta_{12}^0)
-\gamma\sqrt{2}y\right].
\label{s13}
\end{equation}
Expression for non-zero $\theta_{13}$ in terms of the parameters of our model viz. $\epsilon$, the vacuum expectation values $v$ and the quartic couplings $\lambda_i$, ($i=1,2,3$), can be obtained with help of 
Eqs. (\ref{id2}), (\ref{grho}) and (\ref{s13}).

The shift of $\theta_{23}$ from $\pi/4$ can be found from Eq.(\ref{ket3}) as
\begin{equation}
\tan\varphi\equiv\tan(\theta_{23}-\pi/4)=\xi \epsilon.
\label{atmmix}
\end{equation}
The first-order corrections to the first and second ket will contribute to 
changes in $\theta_{12}$. Defining:
\begin{equation}
\beta\equiv \frac{\left[ \frac{y}{\sqrt{2}}\cos 2\theta_{12}^0
+\frac{1}{2}(x-\frac{x'}{2})\sin 2\theta_{12}^0
\right]}{\rho}
\label{beta}
\end{equation}
will lead to corrected solar mixing angle given by,
\begin{equation}
\tan\theta_{12}=\frac{\sin\theta_{12}^0+\epsilon \beta \cos\theta_{12}^0}{\cos\theta_{12}^0-\epsilon \beta \sin\theta_{12}^0}.
\label{solmix}
\end{equation}
Needless to mention, expressions for corrected $\theta_{12}$ in Eq. (\ref{solmix}) and deviations of $\theta_{23}$ from
maximal mixing in Eq. (\ref{atmmix}) can be translated in terms of parameters of this $S3\times Z_2$ symmetric model by applying Eqs. (\ref{id2}), (\ref{grho}), (\ref{xi}) and (\ref{beta}).

In our entire analysis, we have taken $r_{\alpha \beta}$, ($\alpha,\beta=1,2$),
to be real therefore allowing no CP-violation. But one can associate Majorana phases
to masses of the right-handed neutrinos, thus $r_{\alpha \beta}$ can be complex quantities.
Therefore $\epsilon$ can also be complex that can give rise to CP-violation from Eq. (\ref{ket3}).

\begin{figure}[tbh]
\begin{center}
\includegraphics[scale=0.22,angle=0]{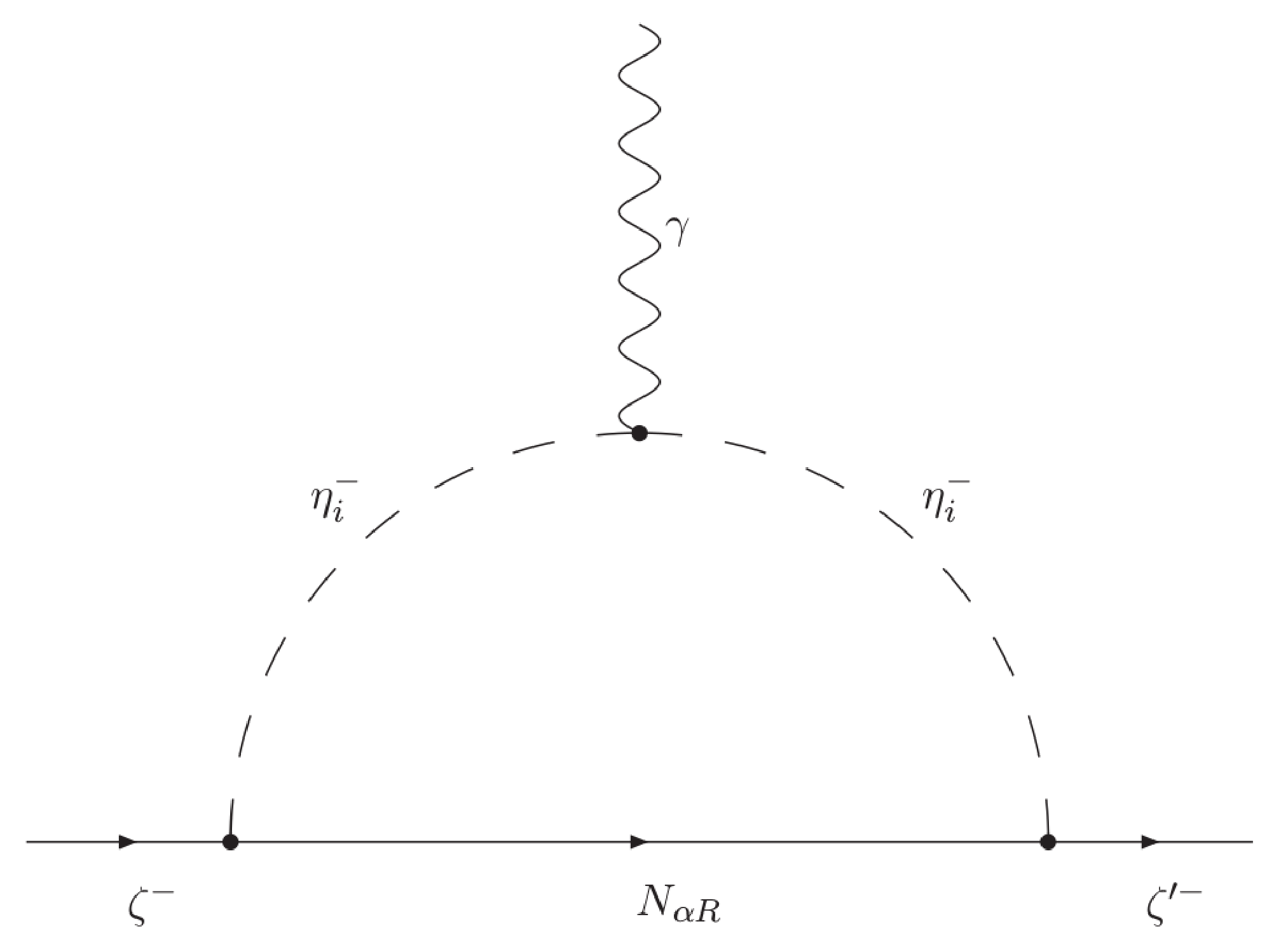}
\end{center}
\caption{\sf  Decays of the charged leptons at one-loop. Here $\zeta^-$ and $\zeta'^-$ stands for
$(e^-,\mu^-,\tau^-)$. For charged lepton flavour violating (LFV) processes $\zeta^- \ne \zeta'^-$.
Kinematically, only $\mu^-\rightarrow e^-\gamma$, 
$\tau^-\rightarrow e^-\gamma$ and $\tau^-\rightarrow \mu^-\gamma$ are allowed 
and are therefore searched for. $S3$ symmetry forbids LFV processes at one-loop 
level in this model.
}
\label{lfvs3} 
\end{figure}

Finally, we want to make a remark on the flavour changing decays of the
charged leptons.
For charged lepton flavour violation (LFV) one requires the part of the Yukawa
Lagrangian similar to Eq. (\ref{yukawa}):
\begin{equation}
\mathscr{L}_{{\rm LFV}}= y_1\left[(\overline{N}_{2R}\eta_2^++\overline{N}_{1R}\eta_1^+)e^-\right]
+y_2\left[(\overline{N}_{1R}\eta_2^+)\tau^-+ (\overline{N}_{2R}\eta_1^+)\mu^-\right]+h.c.
\label{lfv}
\end{equation}
At one-loop level LFV processes can take place through diagrams as shown in Fig. \ref{lfvs3}.
From Eq. (\ref{lfv}) it is readily seen that  the $\mu^-\rightarrow e^-\gamma$, 
$\tau^-\rightarrow e^-\gamma$ and $\tau^-\rightarrow \mu^-\gamma$ processes in Fig. \ref{lfvs3} are disallowed in the model. Specifically, the $\eta_i$ and $N_\alpha$ fields needed at the two Yukawa vertices in Fig. \ref{lfvs3}
for these LFV processes to occur can never be matched taking into account Eq. (\ref{lfv}).
Thus these LFV processes are identically zero at one-loop level as long as $S3$ symmetry is conserved.
%---------------------------------------------------------------------------
%---------------------------------------------------------------------------

\section{Conclusion}
In a nutshell, a radiative $S3\times Z_2$ symmetric scheme of scotogenic generation of
realistic neutrino mixing is put forward. The model has two right-handed neutrinos, $N_{1R}$ and $N_{2R}$,
which when maximally mixed can radiatively yield the form of left-handed Majorana neutrino mass 
matrix at one-loop characterized by $\theta_{13}=0$, $\theta_{23}=\pi/4$ and $\theta_{12}^0$
of any of the values specific to the tribimaximal (TBM), Bimaximal (BM)
and Golden Ratio (GR) mixing collectively termed as popular lepton mixings. 
Small deviation from maximal mixing between the two right-handed neutrino states
can produce realistic mixing angles i.e., non-zero $\theta_{13}$, shifts of the atmospheric mixing angle
$\theta_{23}$ from $\pi/4$ and small corrections to $\theta_{12}$.
There are two inert $SU(2)_L$ doublet scalar fields $\eta_i$, ($i=1,2$) in the model.
Since the $\eta_i$ are odd under the action of the unbroken $Z_2$, the lightest among
these two scalars can serve as dark matter.

{\bf Acknowledgements:} 
My sincere thanks to 
Prof. Amitava Raychaudhuri for discussions and valuable suggestions.

%%%%%%%%%%%%%%%%%%%%%%%%%%%%
%%%%%%%%%%%%%%%%%%%%%%%%%%%%%%%%%%%% Acknowledgements %%%%%%%%%%%%%%%%%%%%%%%%%%%%%%%%%%%%%%%%%%
\renewcommand{\thesection}{\Alph{section}} 
\setcounter{section}{0} 
\renewcommand{\theequation}{\thesection.\arabic{equation}}

\setcounter{equation}{0}

\section{Appendix: The group $S3$ }\label{GroupS3}

It is the permutation group of three objects \cite{s3group} and
therefore has $3!=6$ elements. $S3$ has two generators
$A$ and $B$ that satisfy $A^2 = I = B^3$ and $(AB) ~(AB) =
I$.
The group  properties can be clearly understood from
the group table shown in Table \ref{tabS3}.

%-----------------------------------------------------------------------
\begin{table}[tbh]
\begin{center}
\begin{tabular}{|c|c|c|c|c|c|c|}
\hline
& $I$ & $A$  & $B$ & $C$ & $D$  & $F$ \\ \hline 
$I$ & $I$ & $A$  & $B$ & $C$ & $D$  & $F$ \\ \hline
$A$ &  $A$  & $I$ & $C$ & $B$  & $F$ & $D$ \\ \hline
$F$ &  $F$  & $C$ & $I$ & $D$  & $A$ & $B$  \\ \hline
$C$ &  $C$  & $F$ & $D$ & $I$  & $B$ & $A$ \\ \hline
$D$ &  $D$  & $B$ & $A$ & $F$  & $I$ & $C$ \\ \hline
$B$ &  $B$  & $D$ & $F$ & $A$  & $C$ & $I$  \\ \hline
\end{tabular}
\end{center}
\caption{\sf{ The group table of the discrete symmetry $S3$.}}
\label{tabS3}
\end{table}
%-----------------------------------------------------------------------

It has two one-dimensional representations $1$
and $1'$, as well as one two-dimensional representation $2$.
The one dimensional representation $1$ is immune to both
$A$ and $B$ whereas $1^{\prime}$ flips sign when acted by $A$.
In two-dimension, the group can be represented by the following 
matrices that obey all the properties discussed so far:
\begin{equation}
I = \pmatrix{1 & 0 \cr 0 & 1}\;\;,\;\; A = \pmatrix{0 & 1 \cr 1 &
0}\;\;,\;\; B = \pmatrix{\omega & 0 \cr 0 & \omega^2} \;\;.
\label{S3_21}
\end{equation}
Here $\omega = e^{2\pi i/3}$ is a cube root of one.
With the generators in Eq. (\ref{S3_21}), we can construct 
the rest of the members of the group as:
\begin{equation}
C = \pmatrix{0 & \omega^2 \cr \omega & 0}\;\;,\;\; D = \pmatrix{0
& \omega \cr \omega^2 & 0}\;\;,\;\; F = \pmatrix{\omega^2 & 0 \cr 0
& \omega} \;\;.
\label{S3_22}
\end{equation}
$S3$ is characterized by the following product rules,
\begin{equation}
1 \times 1^\prime = 1^\prime, ~1^\prime \times 1^\prime = 1,
~{\rm and} ~2 \times 2 = 2 + 1 + 1^\prime \;\;.
\label{S3prodapp}
\end{equation}
All the matrices $M_{ij}$ in Eqs. (\ref{S3_21}) and (\ref{S3_22}) obey,
\begin{equation}
\sum_{j,l ~= 1,2} \alpha_{jl} ~M_{ij} ~M_{kl} = \alpha_{ik} \;\;.
\label{s3inv}
\end{equation}
Here $\alpha_{ij} = 0$ if $i = j$ and $\alpha_{ij} = 1$ if $i \neq j$. 

Let $\Phi \equiv \pmatrix {\phi_1 \cr \phi_2}$ and $\Psi \equiv
\pmatrix {\psi_1 \cr \psi_2}$  be two doublets of $S3$ which when 
combined according to Eq. (\ref{S3prodapp}) will yield:
\begin{equation} 
\phi_1 \psi_2 + \phi_2 \psi_1 \equiv 1 \;\; ,
\;\; \phi_1 \psi_2 - \phi_2 \psi_1 \equiv 1^\prime \;\; {\rm
and} \;\; \pmatrix{\phi_2 \psi_2 \cr \phi_1 \psi_1} \equiv 2 \;\;.
\label{S3_prod}
\end{equation}
Often, we have to work with Hermitian conjugate of the fields.
Owing to the properties of the complex representations of $S3$, [say,
as for $B$ displayed in Eq. (\ref{S3_21})], the hermitian
conjugate of $\Phi$ is given by $\Phi^\dagger
\equiv \pmatrix {\phi_2^\dagger \cr \phi_1^\dagger}$. This
$\Phi^\dagger$ when combined with $\Psi$, keeping Eq. (\ref{S3prodapp})
in mind, we get,
\begin{equation} 
\phi_2^\dagger \psi_2 + \phi_1^\dagger \psi_1 \equiv 1 \;\; ,
\;\; \phi_2^\dagger \psi_2 - \phi_1^\dagger \psi_1 \equiv 1^\prime \;\; {\rm
and} \;\; \pmatrix{\phi_1^\dagger \psi_2 \cr \phi_2^\dagger
\psi_1} \equiv 2 \;\;.
\label{S3_prod2}
\end{equation}
Eqs. (\ref{S3_prod}) and (\ref{S3_prod2}) play a 
pivotal role in determining the structure of the mass matrices
in the model. 

\setcounter{equation}{0}
\section{Appendix: The scalar potential}\label{potential}
The scalar sector of the model as can be seen from Table. \ref{fields}, 
comprises of two inert $SU(2)_L$ doublets, 
$\eta_i\equiv(\eta_i^+ \eta_i^0)^T$, ($i=1,2$), forming a doublet under $S3$
denoted by $\eta$ and two other $SU(2)_L$ doublet scalar fields 
$\Phi_j\equiv(\phi_j^+ \phi_j^0)^T$, ($j=1,2$), represented by $\Phi$, transforming
as a doublet under $S3$.
Under the unbroken $Z_2$, $\eta$ is odd whereas $\Phi$ is even. 
Thus after SSB, $\phi_j^0$ can acquire vevs $v_j$,($j=1,2$), but the 
$\eta_i^0$ cannot. 
The complete scalar potential consisting
of all the terms allowed by the SM gauge symmetry and $S3\times Z_2$ is given by:
\begin{eqnarray}
V_{total}&=& m^2_\eta \left(\eta_2^\dagger\eta_2+\eta_1^\dagger\eta_1\right)
+m^2_\phi \left(\phi_2^\dagger\phi_2+\phi_1^\dagger\phi_1\right)\nonumber\\
&+& \widetilde{\lambda}_1\left(\eta_2^\dagger\eta_2+\eta_1^\dagger\eta_1\right)^2
+\widetilde{\lambda}_2\left(\eta_2^\dagger\eta_2-\eta_1^\dagger\eta_1\right)^2
+ \widetilde{\lambda}_3 \left(\phi_2^\dagger\phi_2+\phi_1^\dagger\phi_1\right)^2
+\widetilde{\lambda}_4 \left(\phi_2^\dagger\phi_2-\phi_1^\dagger\phi_1\right)^2\nonumber\\
&+&\widetilde{\lambda}_5\left[\left(\eta_2^\dagger\eta_2+\eta_1^\dagger\eta_1\right)
\left(\phi_2^\dagger\phi_2+\phi_1^\dagger\phi_1\right)\right]
+\widetilde{\lambda}_6\left[\left(\eta_2^\dagger\eta_2-\eta_1^\dagger\eta_1\right)
\left(\phi_2^\dagger\phi_2-\phi_1^\dagger\phi_1\right)\right]\nonumber\\
&+&\widetilde{\lambda}_{7}\left[\left( \phi_1^\dagger\phi_2\right)\left( \phi_2^\dagger\phi_1\right)\right]
+\widetilde{\lambda}_{8}\left[\left( \eta_1^\dagger\eta_2\right)\left( \eta_2^\dagger\eta_1\right)\right]\nonumber\\
&+&\widetilde{\lambda}_{9}\left[\left\{\left( \phi_1^\dagger\phi_2\right)\left( \eta_2^\dagger\eta_1\right)\right  \} 
+ \left\{\left( \phi_2^\dagger\phi_1\right)\left( \eta_1^\dagger\eta_2\right)\right  \} 
\right] + V_{relevant} 
\label{totalpotential}
\end{eqnarray}
where,
\begin{eqnarray}
V_{relevant}&=& \lambda_1 \left[ \left\{(\eta_2^\dagger \phi_2+\eta_1^\dagger \phi_1)^2 \right  \}  + h.c.\right]+ \lambda_2\left[ \left\{(\eta_2^\dagger \phi_2- \eta_1^\dagger \phi_1 )^2 \right \} +h.c.
\right]\nonumber\\
&+&\lambda_3\left[ \left\{(\eta_1^\dagger \phi_2)(\eta_2^\dagger \phi_1)+(\eta_2^\dagger \phi_1)(\eta_1^\dagger \phi_2)\right  \} 
+h.c.\right].
\label{potential2}
\end{eqnarray}
Since at the four-point scalar vertex in Fig. \ref{radfig}, two $\phi$ 
are destroyed and two $\eta$ are created, the terms only of $(\eta^\dagger \phi)(\eta^\dagger \phi)$
type play a crucial role in determining the neutrino mass matrix.
Thus we call these terms as the relevant part of the scalar potential, represented by 
$V_{relevant}$ in Eq. (\ref{potential2}).
The quartic couplings $\lambda_j$ ($j=1,2,3$) appearing in Eq. (\ref{potential2})
were taken to be real for the analysis.

\end{document}